\documentclass[aps,jcp,superscriptaddress,amsmath,amssymb,twocolumn]{revtex4-2} 

\usepackage{chemformula} 
\usepackage[T1]{fontenc} 
\usepackage{braket} 
\usepackage{siunitx}
\usepackage[version=4]{mhchem}
\usepackage{hyperref} 

\newcommand{\be}{\begin{equation}}
\newcommand{\ee}{\end{equation}}
\newcommand{\bea}{\begin{eqnarray}}
\newcommand{\eea}{\end{eqnarray}}

\newcommand{\eqa}{\begin{equation}}
\newcommand{\eqz}{\end{equation}}
\newcommand{\eqma}{\begin{eqnarray}}
\newcommand{\eqmz}{\end{eqnarray}}

\newcommand{\bos}[1]{\boldsymbol{#1}}

\usepackage{array}
\newcolumntype{R}[1]{>{\raggedleft  \arraybackslash}p{#1}@{} }
\newcolumntype{C}[1]{>{\centering \arraybackslash}p{#1}@{} }

\newcommand{\cm}{$\text{cm}^{-1}$}

\newcommand{\nEpv}{$E_\mathrm{PV}$}
\newcommand{\nEPV}{$E_\mathrm{PV}$}
\newcommand{\Epv}{$\Delta E_\mathrm{PV}$}
\newcommand{\EPV}{$\Delta E_\mathrm{PV}$}

\newcommand{\mr}[1]{\mathrm{#1}}
\newcommand{\delpv}{$\Delta\nu_\mr{PV}$}

\keywords{
parity violation, 
large-amplitude motion, 
relativistic calculation,
exact quantum dynamics,
GENIUSH,
DIRAC,
}

\begin{document}
\title[1D PV]
  {Strong parity-violation effects induced by large-amplitude motions: \\
  A quantum-dynamics study of substituted chiral methanols }

\author{Ayaki Sunaga}
\affiliation{%
ELTE, E\"otv\"os Lor\'and University, Institute of Chemistry, P\'azm\'any P\'eter s\'et\'any 1/A 1117 Budapest, Hungary
}%
\email{ayaki.sunaga@ttk.elte.hu,sunagaayaki@gmail.com}


\begin{abstract}

An enhanced mechanism is proposed for the large-amplitude-motion-induced parity-violating frequency by integrating the exact quantum dynamics method with the relativistic electronic structure theory. 
The torsional wavefunctions and PV frequency shifts are obtained by the exact quantum dynamics method. 
The potential energy curve and PV energy along the torsional coordinates are calculated using the extended atomic mean-field two-component Hamiltonian. 
 The predicted PV frequency shift for the torsional transition of CFClBrOH is approximately 100 times larger than that of the conventional C-F stretching mode of CHFClBr. The maximum PV frequency shift (3.2 Hz) is obtained in the CHBrIOH molecule.

\end{abstract}
\maketitle

\section{Introduction}
What will humans discover after improving the precision of molecular spectra? 
The advancement of cooling techniques for polyatomic molecules \cite{Miyamoto2022CommunChem,Aiello2022NatCommun_C2H2,Anderegg2023Science_CaOH,Vilas2024Nature_tweezer_CaOH} can transcend the precision of traditional spectroscopies.
Although the rotation-vibrational spectra are mainly generated by electromagnetic interactions, the parity-violating (PV) interaction may split them.
The PV interaction was first observed in nuclear beta decay \cite{Wu1957PR_beta}, using the optical-rotation technique \cite{Barkov1978JETPLett_Bi_PV,Barkov1979PLB_Bi_PV} as well as in atomic forbidden transition due to a nuclear anapole moment \cite{Wood1997Science_anapole}. 
Various molecules have been investigated as will be discussed. However, only the upper limit of the PV frequency shift \delpv\ has been reported using a methane-like chiral molecule, CHFClBr \cite{Daussy1999PRL_CHFClBr,Ziskind2002EPJD_CHFClBr}.
Although the PV energy difference between the enantiomer (\Epv) can appear in the framework of the standard model \cite{Berger2004_PV,Quack2008ARPC,Berger2019_Parity_violation}, the molecular PV shift may be a probe of cold dark matter \cite{Bargueno2008EPL_PV_DM,Gaul2020PRL_PV_DM}.


As \nEpv\ increases rapidly with an increase in the nuclear charge ($Z$) becomes large \cite{ZelDovich1977PZETF_Z5,Rein1979PLA,Hegstrom1980JCP,Laerdahl1999PRA,Bast2011PCCP} (initially estimated for an atomic case \cite{Bouchiat1974JP}),
a naive strategy for enhancing \Epv\ involves incorporating the heaviest possible element into the molecule (e.g., tungsten \cite{Figgen2010JCP_W_PV}, rhenium \cite{Schwerdtfeger2003ACIE,Saleh2018Chirality_Re_PV,Sahu2023arXiv_Re_PV}, osmium \cite{Fiechter2022JPCL_Os_PV}, iridium and bismuth \cite{Schwerdtfeger2003ACIE}, astatine \cite{Berger2007MP_CHFClBr}, uranium \cite{Wormit2014PCCP}).
In practice, however, there is a limit to how heavy an element can be as long as $Z\leq118$; moreover, some heavy elements are radioactive. 
Another strategy, such as finding a 
PV-effect-sensitive molecular species, is significantly important. The enhancement of PV shifts in the vibrational shifts in methane-like chiral cations (e.g., \delpv\ = 1.8 Hz in \ce{CHDBrI+}) has been recently reported \cite{Eduardus2023CC}.

Large \nEpv\ variations in large-amplitude motion (LAM) are well-known phenomena, although they are rarely applied to PV shift enhancement.
For example, in \ce{H2X2} (where X = O, S, Se, Te, Po) molecules, \nEPV\ as a function of the dihedral angle ($\theta$) behaves as a sinusoidal curve ($\sim \sin(2\theta)$)  \cite{Lazzeretti1997CPL_H2O2,Bakasov1998JCP_H2O2,Laerdahl1999PRA,Thyssen2000PRL_CCSD_T,Berger2005JCP_H2X2_method,Bast2011PCCP}, indicating that \nEpv\ exhibits maximum and minimum values with different signs in $0^\circ \leq \theta\leq180^\circ$.
Therefore, is it possible to design molecules whose initial and final transition states take the minima and maxima ranges of \nEPV?

To exploit this fascinating property of the \nEPV\ curve, the torsional vibrational motion, 
which cannot be described by the conventional harmonic approximation, must be explored. An ideal scenario is that two torsional wavefunctions of the target transition are located over a torsional-angle region with different \nEpv\ values. However, the \nEPV\ values are relatively similar for a torsional transition in the case of tetratomic molecules, such as ClSSCl (Figure 2 of Ref.~ \citenum{Berger2001ACIE}) and HOSH (Fig. 2 of Ref.~ \citenum{Quack2003HCA}). 
Another example of torsional motion is the internal rotation of the methyl unit. Notably,
Berger $et$ $al.$ reported the one-dimensional (1D) calculation of a methanol isotopomer (HDTCOH), although the maximum PV shift was approximately $6 \times 10^{-5}$ Hz \cite{Berger2003HCA_CHDTOH}. A heavier molecule, including a torsional motion (GeClBrI-\ce{CF3}), was calculated \cite{Schwerdtfeger2003ACIE}, although it is fixed at the optimized structure, and its torsional motion was not investigated.


Here, the PV shift in the torsional motions of the CXYZ unit in CXYZOH (X, Y, Z = H, F, Cl, Br, I) molecules was investigated. Strong PV enhancements are observed due to the localization of the torsional wavefunction to the geometries exhibiting opposite \Epv\ signs. The results obtained for CHBrIOH, CFBrIOH, CFClBrOH, and CClBrIOH molecules are reported as examples, and those for the other molecular species are reported in the Supplementary Material (SM).

\begin{figure*}[!htbp]
    \centering
    \includegraphics[width=1.0\linewidth]{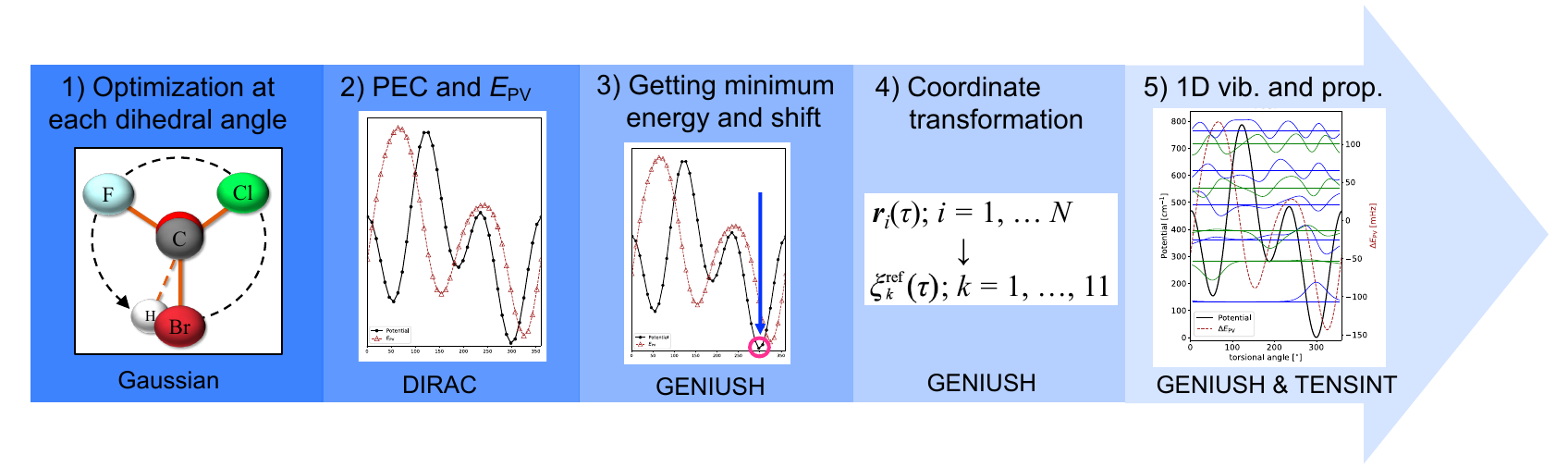}
    \caption{Overview of the steps for obtaining \delpv\ for the torsional states of CXYZOH molecules and utilized programs. From the left, 1-2) the electronic structure calculations, 3) energy shift, 4) coordinate transformations, and 5) quantum dynamics calculations are performed. The figure for step 1) shows the direction of the internal rotation of the CXYZ unit and the geometry with $\tau_1\approx 10$.}
    \label{fig1:overview}
\end{figure*}

\section{Theory}
Figure \ref{fig1:overview} shows the flow diagram for obtaining the PV frequency of the torsional states.
%
%

\begin{figure}[!htbp]
    \centering
    \includegraphics[width=0.6
    \linewidth]{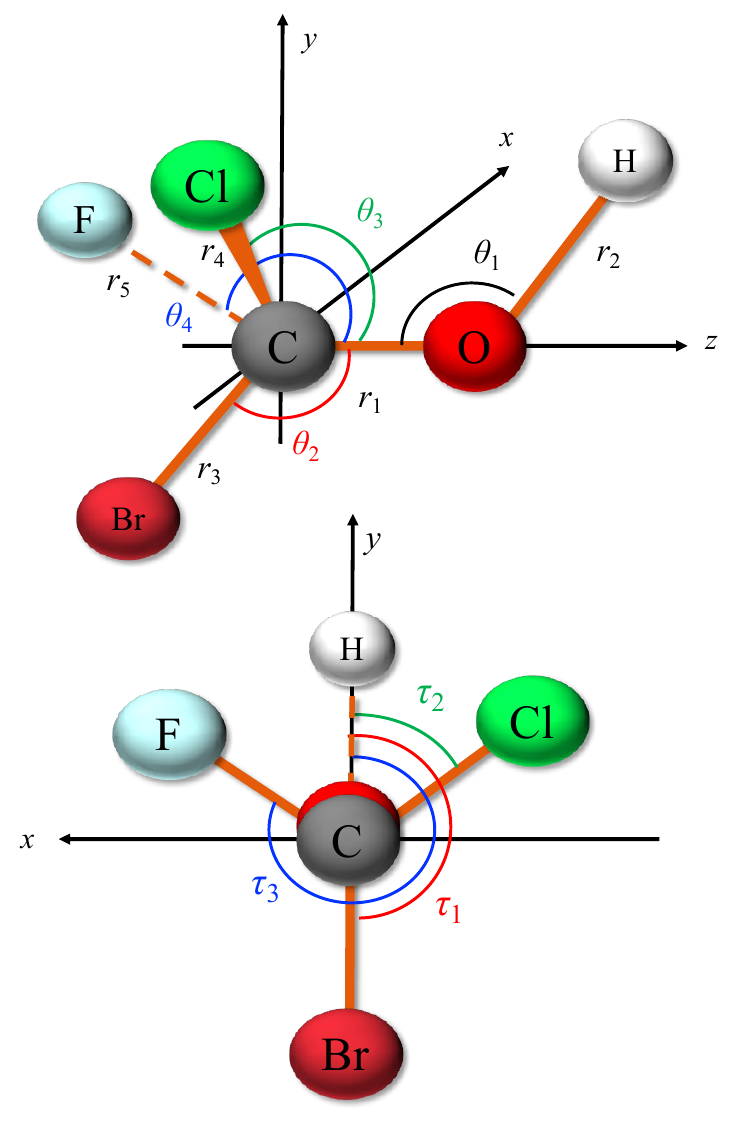}
    \caption{Visualization of a CXYZOH molecule ($S$-CFClBrOH as an example) and the primitive internal coordinates in the right-handed body-fixed Cartesian frame. In the configuration of this figure, $\tau_1=180^\circ$.}
    \label{fig2:coord}
\end{figure}

\begin{enumerate}
\item Geometry optimization was performed at each X-C-O-H dihedral angle ($\tau_1$ in Figure \ref{fig2:coord}). The calculations were performed over an equidistant grid of $\tau$ points, $\tau_{1,n}=(n-1)9^\text{o}\ (n=1,\ldots,41)$. Figure \ref{fig3:relax} shows the direction of the internal rotation; CFClBrOH was used as an example.
\item The total energies (the potential energy curve, PEC) and \EPV\ were calculated using optimized geometries at $\tau_{1,n}$ $(n=1,\ldots,41$; solid black line in Figure \ref{fig3:relax}).
\item The minimum energy on the PEC was obtained using the simplex code \cite{sunaga2024JCTC}, and the PEC was shifted so that its bottom could become zero \cm.
\item The Cartesian coordinates optimized in Step 1 with respect to $\tau_1$ were transformed into the relaxed reference internal coordinates. 
First, we extracted $\tau_1,\tau_2,$ and $\tau_3$ from the Cartesian coordinate defined in Eq. \ref{eq:cart_sym} and obtain the torsional angle ($\tau$; Eq.~\ref{eq:lin_com}). 
Further, $\xi^\mr{ref}_k(\tau_n)$ ($n=1,2,...,37$, which covers $0^\circ\leq\tau\leq360^\circ$) is obtained from $\bos{r}^\mr{ref}_i(\tau_n)$, 
and these discrete coordinates $\xi^\mr{ref}_k(\tau_n)$ are interpolated to obtain $\xi^\mr{ref}_k(\tau)$. 
Additionally, the horizontal axis of the potential energy is also transformed ($V(\tau_{1,n}) \rightarrow V(\tau_{n}))$, while $V(\tau_{n})$ is interpolated.
\item A 1D vibration calculations (considering only the torsional motion) was performed, and \delpv\ at each torsional state was obtained using the expectation value with respect to the vibrational wavefunction (Figure \ref{fig4:vib_pv} and Table \ref{tbl:assign})
\end{enumerate}

%
%
The following definition was used to obtain the body-fixed Cartesian coordinate (Figure \ref{fig2:coord}), following Ref. \citenum{sunaga2024JCTC} 

\begin{align}\label{eq:cart_sym}
&\boldsymbol{r}_{\mathrm{C}}=  \mathbf{0}, \quad \boldsymbol{r}_{\mathrm{O}}=\left(\begin{array}{c}0 \\ 0 \\ r_1\end{array}\right), \\ \nonumber
&\boldsymbol{r}_{\mathrm{H}}=  \boldsymbol{r}_{\mathrm{O}}+\left(\begin{array}{c}0 \\ r_2 \cos \left(\theta_1-\pi / 2\right) \\ r_2 \sin \left(\theta_1-\pi / 2\right)\end{array}\right), \\ \nonumber
&\boldsymbol{r}_\mr{Z}=  \left(\begin{array}{c}-r_3 \sin \theta_2 \sin \tau_1 \\ r_3 \sin \theta_2 \cos \tau_1 \\ r_3 \cos \theta_2\end{array}\right),   \quad 
\boldsymbol{r}_\mr{Y}=  \left(\begin{array}{c}-r_4 \sin \theta_3 \sin \tau_2 \\ r_4 \sin \theta_3 \cos \tau_2 \\ r_4 \cos \theta_3\end{array}\right), \\ \nonumber
&\boldsymbol{r}_\mr{X}=  \left(\begin{array}{c}-r_5 \sin \theta_4 \sin \tau_3 \\ r_5 \sin \theta_4 \cos \tau_3 \\ r_5 \cos \theta_4\end{array}\right) ,
\end{align}
where the atomic number increases from atoms X-Z. 
The frame for the internal rotation of the CXYZ unit was also defined based on the body-fixed Cartesian coordinate definition (Eq.~\ref{eq:cart_sym}). 
Furthermore, the center of mass was fixed at the origin of the body-fixed coordinate system, which offers an additional constant shift vector to each vector in Eq.~\ref{eq:cart_sym}.
 These coordinate definitions were used in the vibrational calculations to prevent the rotation of the frame. 

%
%
The electronic structure calculations, including geometry optimization, PEC, and \Epv\ calculations, were performed on structures parametrized by H-C-O-Z dihedral angle, $\tau_1$, 
whereas the torsional motion in the quantum dynamics calculation was described by $\tau$, defined as a linear combination of $\tau_1,\tau_2,$ and $\tau_3$
\begin{align} \label{eq:lin_com}
 &\tau= \frac{1}{3}\left(\tau_1+\tau_2+\tau_3\right), \\ \nonumber
 &\varphi_{1}=\frac{1} {\sqrt{2}}\left(\tau_2-\tau_3\right), \quad 
 \quad\varphi_{2}=\frac{1} {\sqrt{6}}\left(2 \tau_1-\tau_2-\tau_3\right) \; .
\end{align}
This definition facilitated the identification of $\tau$ as the single LAM in the system, with the $\varphi_1,\varphi_2$ small-amplitude motions (SAMs). This extraction technique for $\tau$ was initially proposed in Refs. \cite{Meyer1969JCP_tau,Bell1994JMS_tau}. 
For subsequent convenience, we define the following compact notation for the internal coordinates,
\be
  \boldsymbol{\rho} 
  =
  \left(\boldsymbol{\xi}; \tau \right)=\left(r_1, r_2, r_3, r_4, r_5, \theta_1, \theta_2, \theta_3, \theta_4, \varphi_1, \varphi_2; \tau\right) ,
  \label{eq:rhoxitau}
\ee
where $\boldsymbol{\xi}$ collects the SAMs. The mathematically allowed coordinate ranges are defined as $\tau \in[0, 2\pi)$, and the other coordinates are inactive in this study. 

%
%
\subsection{Vibrational Hamiltonian}
The Hamiltonian used to describe the quantum vibrational motion can be expressed in the Podolsky form \cite{Podolsky1928PR}:
\be\label{eq:H_pod}
\hat{H}^{\mathrm{rv}, \mathrm{P}}=\frac{1}{2} \sum_{k=1}^{D} \sum_{l=1}^{D} \tilde{g}^{-1 / 4} \hat{p}_k \tilde{g}^{1 / 2} G_{k l} \hat{p}_l \tilde{g}^{-1 / 4}+V,
\ee
where $D$ is the number of the vibrational degree of freedom in the molecule, $\hat{p}_k=-i\partial/\partial \rho_k$ is the momentum operator with respect to the internal coordinates, $\tilde{g}=\rm{det}(\bos{g})$, $\bos{G}=\bos{g}^{-1}$, and $V$ is the potential energy depending on the coordinates of the nuclei. 
In our vibrational calculations, we considered only the torsional motion (i.e., $D=1$). 
In the GENIUSH program \cite{Matyus2009JCP,Fabri2011JCP_rot},  the mass-weighted metric tensor, $\bos{g}$, was calculated from the $t$-vectors, as follows:
\be
g_{k l}=\sum_{i=1}^N m_i \boldsymbol{t}_{i k}^{\mathrm{T}} \boldsymbol{t}_{i l},
\ee
with the vibrational $t$-vector:
\be\label{eq:t_vib}
\boldsymbol{t}_{i k}=\frac{\partial \boldsymbol{r}_i}{\partial \rho_k}, \quad k=1,2, \ldots, D,
\ee
and the rotational $t$-vector: 
\be\label{eq:t_rot}
\boldsymbol{t}_{i D+a}=\boldsymbol{e}_a \times \boldsymbol{r}_i, \quad a=1(x), 2(y), 3(z),
\ee
where $\boldsymbol{e}_a$ is the unit vector in the body-fixed Cartesian coordinates.
%
%
In the actual calculations, the $t$-vectors are obtained at the discrete-variable representation (DVR) grid points ($\tau_{\alpha}=1,\ldots,37$), i.e., the matrix elements are constructed based on $\bos{r}_{i\alpha}$ rather than $\bos{r}_{i}$ in eqs~\ref{eq:t_vib} and~\ref{eq:t_rot}. 
The reference internal coordinates for obtaining $\bos{r}_{i\alpha}$ are relaxed at each $\tau_{\alpha}$ $[\boldsymbol{\xi}^{\mr{ref}}(\tau_{\alpha})]$, 
where $\boldsymbol{\xi}^{\mr{ref}}(\tau_{\alpha})$ is the geometry that minimizes $V$ at a given $\tau$, $\tau_{\alpha}$. 
Employing the relaxed coordinates indicates that the moment of inertia in this study depends on $\tau$. The wavefunctions and energies associated with Eq. \ref{eq:H_pod} are obtained in a variational manner.

%
%
\subsection{Parity-violating Hamiltonian}


The PV effective Hamiltonians arise from the coupling between the vector [$V;\psi^{\dagger}(\bos{\alpha},\text{i})\psi)$] and axial-vector [$A;\psi^{\dagger}(\bos{\Sigma},\gamma)\psi)$] terms of the Fermion currents \cite{Moskalev1976SPU,Greiner1996_PV,Berger2004_PV,Bast2011PCCP}. 
Employing indices of an electron (e) and a nucleon [proton (p) and nucleon (n)], $V_\mr{e} - A_\mr{p, n}$ and $A_\mr{e} - V_\mr{p, n}$ contribute to the nuclear spin-dependent term and independent term, respectively. 
The former could appear in nuclear magnetic resonance (NMR) \cite{Barra1986PLA_PV_NMR,Laubender2006PRA_PV_NMR,Weijo2007JCP_PV_NMR_curve,Aucar2022PRA} owing to the contribution from the nuclear spin ($\psi^{\dagger}_\mr{p, n}\bos{\Sigma}\psi_\mr{p, n})$. The latter is our target, where the effective Hamiltonian in the four-component framework is shown as \cite{Moskalev1976SPU,Greiner1996_PV,Berger2004_PV,Bast2011PCCP}
\be\label{eq:H_pv}
\hat{H}_{\mathrm{PV}}=\frac{G_{\mathrm{F}}}{2 \sqrt{2}} \sum_A Q_{W, A} \sum_i \gamma_i^5 \rho_A\left(\mathbf{r}_i-\mathbf{r}_A\right).
\ee
Here, $G_\mr{F}=2.22255\times10^{-14}E_ha^3_0$ is the Fermi coupling constant, $Q_{W, A}=-N_A+Z_A(1-4\sin^2\theta_W)$ is the weak charge of the nucleus $A$, with its neutron number $N_A$ and nuclear charge $Z_A$. The Weinberg mixing angle $\theta_W=0.2319$ \cite{PV_theta_W,Zyla2020PTEP_reiview} was also employed.
$\gamma^5=-i\gamma^0\gamma^1\gamma^2\gamma^3$ is the Dirac matrix that results in the coupling between the large and small components of the wavefunction. Additionally, $\rho_A$ is the charge density of nucleus $A$ for which the Gaussian-type model \cite{visscher1997ADNDT} is employed in this study. 
As the PV interaction is mediated by the heavy $Z_0$ boson, the contact interaction between $\rho_A$ and an electron offers a good approximation. Although Eq. \ref{eq:H_pv} is a four-component expression, it is numerically transformed into a two-component framework (picture-change \cite{Baerends1990JPB_PC}) in this study. 
%
As the sign of the $\hat{H}_\mathrm{PV}$ becomes opposite after parity inversion,
 the energy difference due to the PV interaction between the $R$- and $S$-configuration can be expressed by
\be
 \Delta E_\mr{PV} = 2 E^Z_\mr{PV}; \quad E_{\mathrm{PV}}
=\braket{\Psi^Z_\mr{e}|\hat{H}_{\mathrm{PV}}|\Psi^Z_\mr{e}}.
 \ee
where $Z$ is the target configuration of the molecule ($R$ or $S$).
$E_{\mathrm{PV}}$ is the PV energy obtained from the expectation value with respect to the electronic wavefunction, $\Psi_\mr{e}$.
The PV shift at the vibrational state, $\ket{\Psi_{\nu}}$, is given, as follows:
\be\label{eq:vib_pv}
\Delta \nu_{\mr{PV},\nu} = \braket{\Psi_{\nu}|\Delta E_\mr{PV}(\bos{\rho})|\Psi_{\nu}},
\ee
where $\Delta E_\mr{PV}(\bos{\rho})$ expresses the value of $E_\mr{PV}$ at the nuclear configuration, $\bos{\rho}$. 
We obtained the expected value with respect to the DVR vibrational wavefunction. This indicates that the error due to the truncation of the potential (e.g., the harmonic oscillator and anharmonic correction), which is traditionally employed in this field, is not observed in this manner.


%
%
\section{Computational details}
The vibrational Schr{\"o}dinger equation in a reduced-dimension form was solved using the GENIUSH program \cite{Matyus2009JCP,Fabri2011JCP_rot}, which was recently reviewed in Ref. \citenum{Matyus2023CC}. The derivative in Eq. \ref{eq:t_vib} was obtained numerically (two-point central derivative) using the step size $1.0\times10^{-5}$ bohr. 
Furthermore, 37 Fourier-DVR grid points, which is a sufficient number of grid points as presented in Table S1 of the SM, were used to solve the vibrational Schr\"{o}dinger equation. 
The TENSINT program \cite{Daria2021JCP,Daria2024PCCP} was used to obtain  \delpv\ at each torsional energy level.

Geometry optimization was carried out using the Gaussian16 software package \cite{g16a03}. For the Br and I atoms, we employed the pseudopotentials ECP10MDF and ECP28MDF, respectively, and the cc-pVTZ-PP basis sets \cite{Peterson2003JCP_ccpv-PP_16_17, Peterson2006JPCA_ccpv-PP_17}. For the H, C, O, F, and Cl atoms, we employed the Def2TZVPP basis sets \cite{Weigend2005PCCP_Def2TZVPP}. The PBE1PBE functional \cite{Adamo1999JCP_PBE0,Ernzerhof1999JCP_PBE0} was used for the density functional theory (DFT) functional. Dispersion effects were considered within the semi-empirical D3 van der Waals corrections \cite{Grimme2010JCP_D3}.

To calculate the total electronic energy and \Epv, we employed a development version (git hash: 3970bcf) of the DIRAC program package \cite{saue2020dirac,DIRAC22}. 
We constructed a two-component relativistic Hamiltonian using the exact two-component (X2C) Hamiltonian \cite{Ilias2007JCP} for the one-electron operators and the extended atomic mean-field (eAMF-X2C) approach \cite{Knecht2022JCP_eamfi} to consider the picture-change effect of the two-electron Coulomb and Gaunt terms. The PBE0 functionals \cite{Adamo1999JCP_PBE0,Ernzerhof1999JCP_PBE0} were employed for the electronic structure calculations, and the dyall.3zp basis sets \cite{dyall2016relativistic,Dyall2002TCA_4p5p6p,Dyall2006TCA_4p5p6p} were employed for all the elements in the molecules in the uncontracted form. 
Several tight exponents, which were optimized using the simplex method \cite{SunagaMP2021}, were also added (Table S2 of the SM).
The Gauss-type finite nuclear-charge model \cite{visscher1997ADNDT} was employed for the relativistic calculations.

\section{Results and discussion}

%
%
%
%
\begin{figure*}[!htbp]
    \centering
    \includegraphics[width=0.9\linewidth]{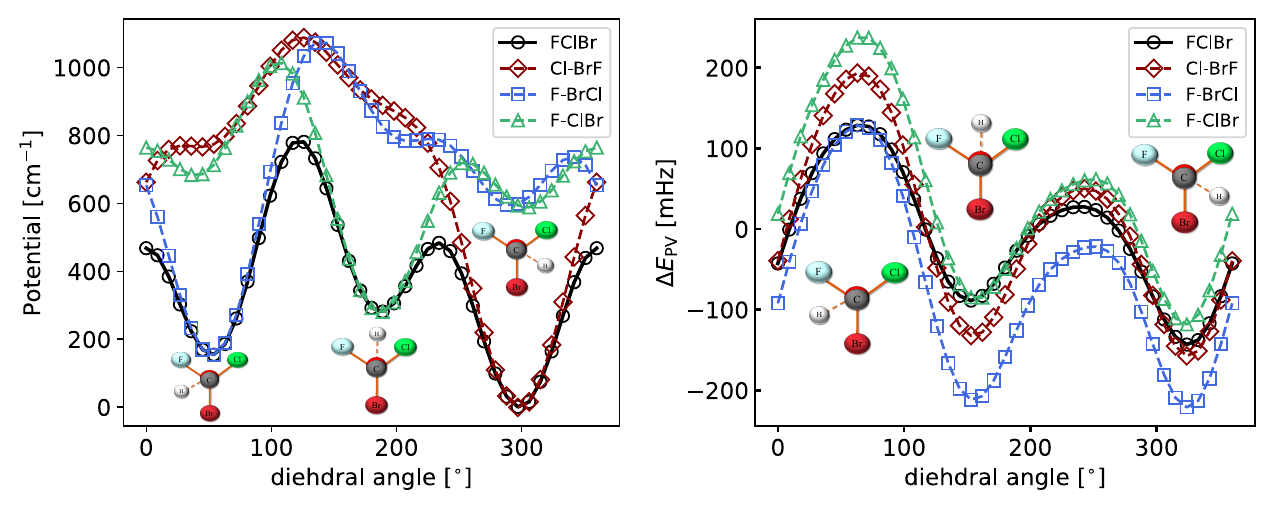}
    \caption{
    PECs (left) and \Epv\ curves (right) of $S$-CFClBrOH obtained under various geometry-optimization conditions.
The solid black line (FClBr) indicates that the geometry optimization was performed at all the grid points (dihedral angle, $\tau_{1,n}$ ($n=1,...,41$). 
The dotted line (legend A-BC) indicates that the geometries were optimized using an initial structure in which the hydrogen of the OH group is positioned between atoms A and B at the PBE1PBE-D3/cc-pVTZ level. The internal coordinates were fixed to generate the PECs.
The molecular figures show the schematic configurations around $\tau_1=60^\circ,180^\circ,300^\circ$.
The PBE0 functional based on the eAMF-X2C Hamiltonian and dyall.3zp basis sets with tight exponents were employed to calculate the PECs and \Epv\ curves.
    }
    \label{fig3:relax}
\end{figure*}

\begin{figure*}
    \centering
    \includegraphics[width=0.9\linewidth]{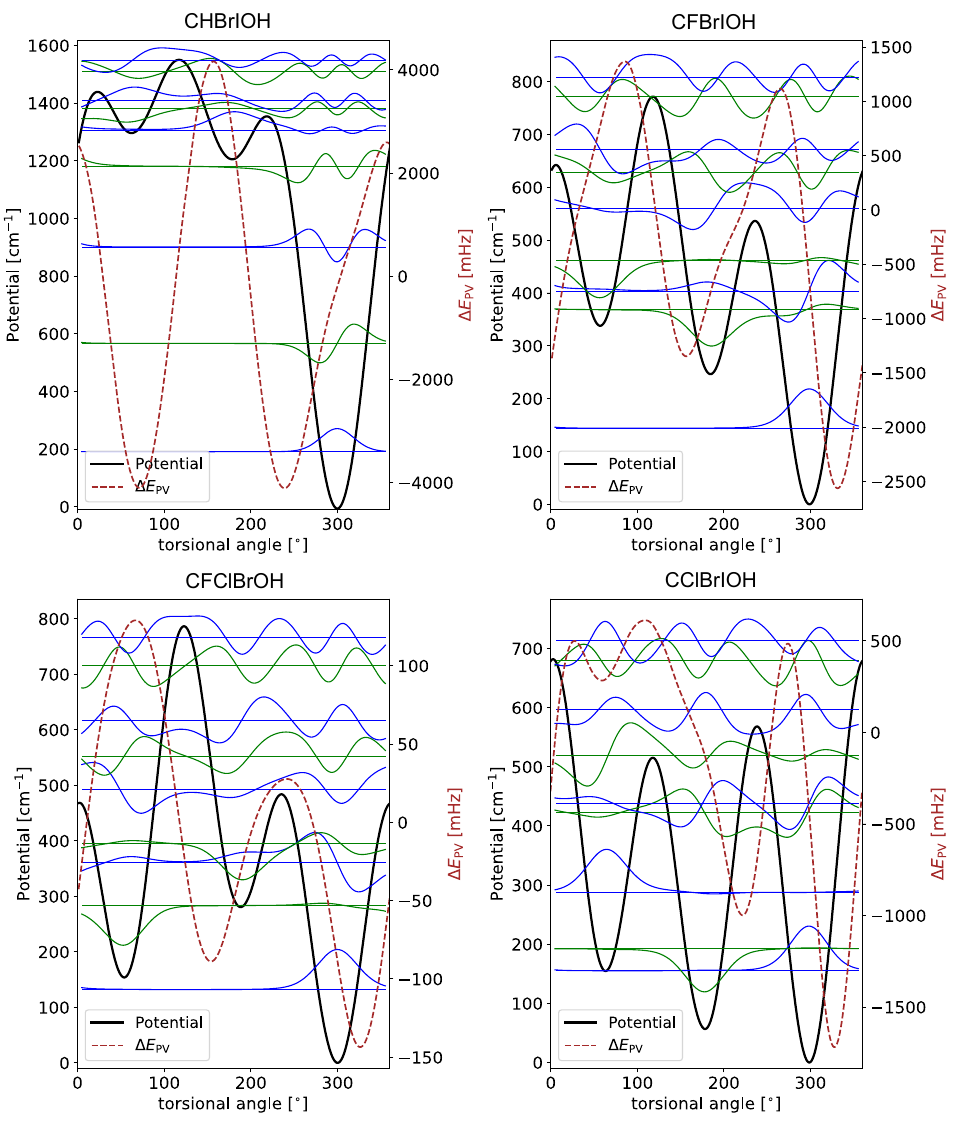}
    \caption{Visualization of the PECs (black solid line; left axis), 1D torsional wavefunctions (left axis), and \Epv\ curves (brown dotted line; right axis) of the $R$-CHBrIOH, $S$-CFBrIOH, $S$-CFClBrOH, and $S$-CClBrIOH molecules. 
    The blue and green solid lines indicate the odd (ground, 3rd, ...) and even-numbered (2nd, 4th, ...) levels of the torsional states. 
The horizontal axis is $\tau$ defined in Eq. \ref{eq:lin_com}. \delpv\ at each torsional level is presented in Table \ref{tbl:assign}.
}
    \label{fig4:vib_pv}
\end{figure*}

Figure \ref{fig3:relax} shows the criticality of the geometry-relaxation effect in the generation of the 1D PEC and \Epv\ curve. 
The fixed-geometry model cannot reproduce the results of the relaxed structure, as some wells in the PEC do not appear in the fixed-geometry model. Although the situation is better in \Epv, 
the fixed-geometry model overestimates the PV sensitivities in many regions. 
From a different perspective, geometries deviating from the optimized torsional path can display larger PV sensitivities, and it 
encourages full-dimensional 
quantum dynamics calculations in future studies. The relaxed structure model was used for the following discussion. Although the shape of the \Epv\ curve appeared complicated, two sinusoidal curves (a local minimum and a local maximum at 0$^\circ$-180$^\circ$ and 180$^\circ$-360$^\circ$, respectively) can be found for some target molecules (see Figures \ref{fig4:vib_pv} and S1-S6 in the SM).


%
%

The enhancement of the PV frequency shift induced by the torsional motion can be visually demonstrated. Figure \ref{fig4:vib_pv} shows the PEC, \Epv\ curve, and torsional wavefunctions of some CXYZOH molecules. 
The \Epv\ curves take various values along $\tau$, 
and the torsional wavefunctions located in different wells can exhibit different  \Epv\ values. 
This yields large \delpv\ values in the transition across the two wells.
This is in contrast to SAMs, where the small displacement from the equilibrium structure contributes to the vibrational wavefunction, and the two states exhibit a similar \delpv\ magnitude. Following Ref.~ \citenum{Eduardus2023CC}, the contribution from the $\partial^2 E_{\mathrm{PV}}(q) /\left.\partial q^2\right|_{q=0} \approx 0$ term is small, yielding a linear-like \Epv\ curve with respect to the vibrational coordinate ($q$). 
The \delpv\ value and vibrational band origin for each torsional state are presented in Table \ref{tbl:assign}. 
Assuming the state control was possible, the maximum sensitivity of 3.2 Hz would be achieved in CHBrIOH. 
The PV frequencies from the ground state exceeded more than 1.0 Hz in the transitions of the CHBrIOH and CFBrIOH molecules. 
%
Compared with the CHFClBr molecule with \delpv\ = 1.867 (--3.781) mHz for the C-F stretching (FCCl bending) fundamental modes \cite{Rauhut2021PRA}, approximately 100- (50-)fold enhancement was observed in CFClBrOH (186.4 mHz) without including a heavy element. The necessary precisions (the ratio between the PV shift and torsional frequency) for detecting the PV shift are $\sim 4\times 10^{-12}$ and $\sim 1\times 10^{-14}$ for CHBrIOH and CFClBrOH, respectively. 

These PV enhancements would not be due to the approximation employed in this study.
The qualitative accuracy of the reduced dimensional models that consider only large-amplitude motion(s) has been confirmed in previous studies~\cite{Senent2001MP_2D,Blasco2003CPL_ch3oh,Csaszar2004JCP,Matyus2009JCP,Ragni2009JPCA_NH3_H3O+_1D}. 
The PBE0 functional offers an acceptable quality (maximum absolute error of 0.5 \text{kcal/mol} from the CCSD(T) level) for the torsional potential energy including halogens (Figure 5 of ref \cite{Tahchieva2018JCTC_tor_DFT}).
The torsional coordinates that offer the maximum, zero, and minimum $E_\mr{PV}$ values do not significantly depend on the methodology choice, e.g., the sinusoidal curves of H$_2$X$_2$ (Fig. 1 of Ref.~\citenum{Bast2011PCCP}), and the torsional curves of dichlorinedioxide and substituted allenes \cite{Horny2015MP_PV}.
Although the maximum and minimum $E_\mr{PV}$ values depend on the methodology, the error due to PBE0 would be much smaller than the enhancement, as the maximum difference between $E_\mr{PV}$ at the PBE0 and CCSD levels of theory is approximately 11\% for the H$_2$X$_2$ (X = O, S, Se, Te, Po) molecules \cite{SunagaMP2021}. The PV enhancement due to the torsional motion was also observed with the CAM-B3LYP* functional \cite{Thierfelder2010PRA_CAMB3LYP*}, as shown in Sec S6 of the SM.

Another advantage of the transitions across the well is that the extremely long lifetime of the torsional excited states is expected. 
When the torsional wavefunctions are mainly localized to one of the three wells, 
they may be recognized as exhibiting different vibrational modes from one that is localized to another well. 
For example, the localized torsional states of \#1, \#2, and \#3 of CClBrIOH can be assigned to hypothetical vibrational modes with different quantum numbers, $\nu_\mr{Cl}$, $\nu_\mr{I}$ and $\nu_\mr{Br}$.
These nearly forbidden transitions produce a long lifetime of the vibrational excited states, resulting in a very narrow line width for the transition.

%
%
\begin{table}[!htbp]
\caption{
Vibrational energies ($\tilde{\nu}$ in \cm) referenced to the zero-point vibrational energy (\#1) and PV frequency shift (\delpv\ in mHz) of the $R$-CHBrIOH, $S$-CFBrIOH, $S$-CFClBrOH, and $S$-CClBrIOH Molecules. $\nu_A$ ($A$ = X, Y, Z) is the quantum number for the well in which the dihedral angle X-C-O-H is close to $180^{\circ}$ at the approximately corresponding $\tau$ values (in degree). The $\nu_A$ represents the torsional quantum number, which is detailed in Sec. S5 of the SM. ``-'' refers to the mixed states that are delocalized to the three wells. 
The recommended transition is highlighted in bold. 
  } \label{tbl:assign}
\begin{tabular}{cccc rr cccc rr}
\hline\hline
$\tau$ & 60 & 180 & 300 & \multicolumn{2}{c}{CHBrIOH} & $\tau$ & 60 & 180 & 300 & \multicolumn{2}{c}{CFBrIOH}\tabularnewline
\# & $\nu_\mr{Br}$ & $\nu_\mr{I}$ & $\nu_\mr{H}$ & $\tilde{\nu}$ & \delpv & \# & $\nu_\mr{Br}$ & $\nu_\mr{I}$ & $\nu_\mr{F}$ & $\tilde{\nu}$ & \delpv \tabularnewline
\hline
1 & 0 & 0 & 1 & 191 & --259.7 & 1 & 0 & 0 & 1 & 144 & \textbf{--674.1}\tabularnewline
2 & 0 & 0 & 2 & 376 & --402.7 & 2 & 0 & 1 & 0 & 224 & --682.2\tabularnewline
3 & 0 & 0 & 3 & 710 & --547.3 & 3 & 0 & 1 & 2 & 260 & --638.1\tabularnewline
4 & 0 & 0 & 4 & 990 & --746.6 & 4 & 1 & 0 & 0 & 317 & \textbf{672.7}\tabularnewline
5 & 0 & 1 & 4 & 1115 & \textbf{1556.9} & 5 & 0 & 2 & 2 & 417 & --483.5\tabularnewline
6 & 1 & 1 & 5 & 1190 & --1277.0 & 6 & 0 & 2 & 3 & 484 & --725.5\tabularnewline
7 & 1 & 1 & 5 & 1218 & \textbf{--1629.4} & 7 & - & - & - & 529 & 204.8\tabularnewline
8 & 1 & 2 & 5 & 1320 & 474.2 & 8 & - & - & - & 629 & --404.7\tabularnewline
9 & - & - & - & 1358 & --578.0 & 9 & - & - & - & 665 & --172.5\tabularnewline
 &  &  &  &  &  &  &  &  &  &  & \tabularnewline
 \hline
$\tau$ & 60 & 180 & 300 & \multicolumn{2}{c}{CFClBrOH} & $\tau$ & 60 & 180 & 300 & \multicolumn{2}{c}{CClBrIOH}\tabularnewline
\# & $\nu_\mr{Cl}$ & $\nu_\mr{Br}$ & $\nu_\mr{F}$ & $\tilde{\nu}$ & \delpv & \# & $\nu_\mr{Br}$ & $\nu_\mr{I}$ & $\nu_\mr{Cl}$ & $\tilde{\nu}$ & \delpv \tabularnewline
\hline
1 & 0 & 0 & 1 & 132 & \textbf{--82.2} & 1 & 0 & 0 & 1 & 155 & \textbf{--479.5}\tabularnewline
2 & 1 & 0 & 0 & 151 & \textbf{104.2} & 2 & 0 & 1 & 0 & 37 & --168.0\tabularnewline
3 & 0 & 0 & 2 & 229 & --54.4 & 3 & 1 & 0 & 0 & 133 & \textbf{367.0}\tabularnewline
4 & 0 & 1 & 2 & 264 & --33.4 & 4 & 0 & 2 & 2 & 267 & --364.4\tabularnewline
5 & 2 & 0 & 3 & 360 & 22.4 & 5 & 0 & 2 & 2 & 284 & --319.1\tabularnewline
6 & - & - & - & 421 & 5.2 & 6 & 2 & 2 & 0 & 364 & 332.7\tabularnewline
7 & - & - & - & 486 & --21.1 & 7 & - & - & - & 442 & --301.3\tabularnewline
8 & - & - & - & 584 & --0.8 & 8 & - & - & - & 524 & --234.6\tabularnewline
9 & - & - & - & 634 & --9.0 & 9 & - & - & - & 559 & --58.1\tabularnewline
\hline\hline
\end{tabular}
\end{table}

%
%

\par

Some torsional transitions appear in the terahertz (THz) range, where lasers are actively developed to achieve stable frequencies \cite{Guo2024APR}, as in the case of quantum cascade lasers \cite{Zeng2020AOM, Vitiello2021APX}. However, the transition to the SAM-LAM coupling mode can reach the wavelengths in the infrared range, where CO$_2$ laser can reach.
In the case of the H-containing species, PV-sensitive transitions are observed in the infrared range, such as the transition between \#1 and \#5 of CHBrIOH presented in Table \ref{tbl:assign}. 
To the best of our knowledge, these substituted methanol molecules have not been synthesized; however, they may be formulated in the gas phase via laser ablation, which is widely employed (e.g., Refs. \citenum{Tokunaga2017NJP,Asselin2017PCCP,Cournol2019QE,Augenbraun2020PRX,Persinger2023PhysRevA}).

%

\section{Conclusions}
In conclusion, an enhancement mechanism for the PV frequency originating in the torsional motions of substituted chiral methanol molecules was demonstrated. 
The torsional transition of the CFClBrOH molecule exhibits approximately 100 times more sensitivity to the PV measurement compared with the conventional C-F stretching motion of the CHFClBr molecule.
The calculated PV shift of CHBrIOH reached 3.2 Hz. 
The torsional energy levels were calculated using the exact quantum dynamics method. The PECs and \Epv\ curves of the torsional motion were calculated using a DFT method based on the two-component relativistic eAMF-X2C Hamiltonian.
The key idea of the enhancement mechanism is that the torsional states located in different wells can have \delpv\ with the opposite sign.

The insights from this study are general and could be combined with other existing ideas, e.g., including heavier elements in the molecule and cationic systems \cite{Eduardus2023CC}. 
A similar enhancement is expected in the PV NMR shielding constant as this property in \ce{H2O2} yields shows a sinusoidal curve \cite{Laubender2006PRA_PV_NMR,Weijo2007JCP_PV_NMR_curve}. 
This study aims to demonstrate the enhancement mechanism of the PV effect. These molecules should be pursued with more accurate calculations, e.g., quantum dynamics calculations taking the full dimension into account, which were demonstrated in methanol \cite{sunaga2024JCTC}, as well as electron-correlated calculations for the PEC and PV curve. Our enhancement mechanism also encourages the development of lasers in the THz range.

\begin{acknowledgments}
The author wishes to acknowledge a financial support from the European Union’s Horizon 2022 research and innovation programme under the Marie Skłodowska-Curie Grant Agreement No. 101105452 (QDMAP) and the JSPS KAKENHI (grant no 21K14643). The author used the computer resource offered under the category of General Projects by the Research Institute for Information Technology, Kyushu University, and the FUJITSU Supercomputer PRIMEHPC FX1000 and FUJITSU Server PRIMERGY GX2570 (Wisteria/BDEC-01) at the Information Technology Center, The University of Tokyo. The author thanks Edit M\'atyus (Budapest) for valuable discussions and Alberto Mart\'in Santa Dar\'ia (Salamanca) for the support in adding the PV module to the TENSINT code.
The author thanks Editage (\url{www.editage.com}) for English language editing.
\end{acknowledgments}

\bibliography{sunaga_0111.bib}

\clearpage

\begin{widetext}
\begin{center}
    \large 
\bfseries{Supporting Information: \\
Strong parity-violation effects induced by large-amplitude motions: \\
  A quantum-dynamics study of substituted chiral methanols}
\end{center}
\end{widetext}


%
%
\section{Employed masses}
In the variational vibration calculation, the following atomic masses were used for the nuclei;  $m_\text{H}=1.007825$~u, $m_\text{C}=12$~u, $m_\text{O}= 15.994915$~u, $m_\text{F}= 18.998403$~u, $m_\text{Cl}= 34.968853$~u, $m_\text{Br}= 78.918330$~u, and $m_\text{I}= 126.90447$~u.


%
%
\section{Convergence with respect to the DVR grid points}
The convergence with respect to the discrete variable representation (DVR) grid points and functions was tested using the CHFIOH molecule (Table \ref{tbl:1D_grid}). We obtained the convergence 0.1 \cm\ for vibrational energies and 0.1 mHz for \delpv\ using 37 DVR grid points up to 20th vibrational states. The 37 DVR grid points and functions are employed for the calculations in the main text.

%
%
\begin{table}[!htbp]
\caption{%
Torsional vibrational energy levels of CHFIOH ($\tilde{\nu}$ in \cm), referenced to the ZPVE (\#1), and parity-violating (PV) frequency ($\Delta\nu^\mr{PV}$ in mHz) for an increasing number of Fourier DVR points and functions, $M_{\tau}$. 
$\delta_{M_{\tau}} = \tilde{\nu}_{M_{\tau} + 2}-\tilde{\nu}_{M_{\tau}}$ in \cm\ for vibrational energies, and $\delta_{M_{\tau}} = \Delta\nu^\mr{PV}_{M_{\tau} + 2}-\Delta\nu^\mr{PV}_{M_{\tau}}$ in mHz for the PV frequencies are also listed.
}
\label{tbl:1D_grid}
\scalebox{0.9}{

\begin{tabular}{@{}c SSSS SSSS@{}}
\hline
 & \multicolumn{4}{c}{$\Tilde{\nu}$} & \multicolumn{4}{c}{$\Delta \nu^\mr{PV}$}\tabularnewline
 \# & $\delta_{31}$ & $\delta_{33}$ & $\delta_{35}$ & 37 & $\delta_{31}$ & $\delta_{33}$ & $\delta_{35}$ & 37\tabularnewline
 \hline
1 & 0.0 & 0.0 & 0.0 & 188.2 & 0.0 & 0.0 & 0.0 & -603.6\tabularnewline
2 & 0.0 & 0.0 & 0.0 & 357.6 & 0.0 & 0.0 & 0.0 & -522.9\tabularnewline
3 & 0.0 & 0.0 & 0.0 & 671.3 & 0.0 & 0.0 & 0.0 & -406.4\tabularnewline
4 & 0.0 & 0.0 & 0.0 & 908.9 & 0.0 & 0.0 & 0.0 & -63.1\tabularnewline
5 & 0.0 & 0.0 & 0.0 & 963.0 & 0.0 & -0.1 & 0.0 & -154.4\tabularnewline
6 & 0.0 & 0.0 & 0.0 & 1103.2 & 0.0 & 0.0 & 0.0 & -125.7\tabularnewline
7 & 0.0 & 0.0 & 0.0 & 1181.1 & 0.0 & 0.0 & 0.0 & -166.4\tabularnewline
8 & 0.0 & 0.0 & 0.0 & 1229.7 & 0.0 & 0.0 & 0.0 & -172.8\tabularnewline
9 & 0.0 & 0.0 & 0.0 & 1292.5 & 0.0 & 0.0 & 0.0 & -225.9\tabularnewline
10 & 0.0 & 0.0 & 0.0 & 1422.9 & 0.1 & 0.0 & 0.0 & -181.4\tabularnewline
11 & 0.0 & 0.0 & 0.0 & 1434.3 & -0.1 & 0.0 & 0.0 & -232.0\tabularnewline
12 & 0.0 & 0.0 & 0.0 & 1645.5 & -0.2 & 0.0 & 0.0 & -202.6\tabularnewline
13 & 0.0 & 0.0 & 0.0 & 1647.1 & 0.2 & 0.0 & 0.0 & -208.7\tabularnewline
14 & 0.0 & 0.0 & 0.0 & 1915.2 & 0.0 & 0.1 & -0.1 & -206.8\tabularnewline
15 & 0.0 & 0.0 & 0.0 & 1915.4 & 0.0 & -0.1 & 0.1 & -207.4\tabularnewline
16 & 0.0 & 0.0 & 0.0 & 2231.7 & -0.2 & 0.0 & 0.0 & -209.3\tabularnewline
17 & 0.0 & 0.0 & 0.0 & 2231.8 & 0.2 & 0.0 & 0.0 & -209.2\tabularnewline
18 & -0.2 & 0.0 & 0.0 & 2594.0 & 0.4 & -0.1 & 0.0 & -211.2\tabularnewline
19 & 0.2 & 0.0 & 0.0 & 2594.1 & -0.3 & 0.1 & 0.0 & -211.4\tabularnewline
20 & -1.1 & -0.1 & 0.0 & 3001.2 & 1.9 & 0.2 & 0.0 & -213.0\tabularnewline
\hline
\end{tabular}
}
\end{table}

%
%
\section{Tight exponents}
The added exponents to the dyall.3zp basis sets are listed in Table \ref{tbl:exponent}. These tight exponents are key to the accurate description of the wavefunction in the core region. 
These exponents were optimized to reproduce the PV matrix elements of the numerical calculation using the GRASP code \cite{dyall1989grasp} within the 1\% error. A more detailed explanation of the optimization procedure is described in ref \cite{SunagaMP2021}. 

\begin{table*}[!htbp]
\caption{%
Existing and optimized additional exponents for the O, F, Cl, Br, and I atoms of the dyall.3zp basis sets. The O's exponents were obtained in Ref. \cite{SunagaMP2021}.
}
\label{tbl:exponent}
\begin{tabular}{l|c|c| cc |c|c}
\hline
 & O & F & \multicolumn{2}{c|}{Cl} & Br & I\tabularnewline
 & $p$ & $p$ & $s$ & $p$ & $p$ & $p$\tabularnewline
 \hline
second tightest & 4.5531279E+01 & 5.8279110E+01 & 7.8714993E+05 & 4.7203082E+02 & 2.4820723E+04 & 1.1662681E+06\tabularnewline
tightest & 1.9430165E+02 & 2.4929585E+02 & 6.4063687E+06 & 2.1571374E+03 & 1.6203213E+05 & 7.5310209E+06\tabularnewline
added & 2.8126178E+03 & 5.1263846E+03 & 5.9958092E+07 & 3.6831954E+04 & 1.8253547E+06 & 5.4139599E+07\tabularnewline
 & 1.9396596E+05 & 1.5303883E+04 &  & 1.3685631E+06 & 2.8240083E+07 & \tabularnewline
 &  & 2.8277597E+04 &  & 2.4578813E+07 &  & \tabularnewline
 \hline
\end{tabular}
\end{table*}

%
%

%
%
\section{PECs, wavefunctions, and PV curves}
All the target molecules are summarized in Table \ref{tbl:config}. The configurations ($R$- or $S$-) were determined so that the deepest PEC well is located around $\tau=300^\circ$.
%
%
\begin{table}[!htbp]
\caption{%
Summary of the target molecules.
}
\label{tbl:config}
\begin{tabular}{ll}
\hline 
$R$-configuration & $S$-configuration \tabularnewline
\hline 
CHFClOH &  CFClBrOH  \tabularnewline
CHFBrOH & CFClIOH  \tabularnewline
CHFIOH &   CFBrIOH  \tabularnewline
CHClBrOH &  CClBrIOH  \tabularnewline
CHClIOH & \tabularnewline
CHBrIOH & \tabularnewline
\hline 
\end{tabular}
\end{table}

Figures \ref{fig:HFCl}-\ref{fig:FClI} display the potential energy curves (PECs), 1D torsional wavefunctions, and parity-violating energy (\Epv) curves of the molecules that are not shown in the main text.
In these figures, the blue and green solid lines indicate the odd (ground, 3rd, ...) and even-numbered (2nd, 4th, ...) torsional state levels. 
The horizontal axis represents the torsional angle ($\tau$) defined in Eq. 2 of the main text. An example of the direction of the internal rotation of the CXYZ unit is shown in FIG. 3 of the main text. The atomic number increases from atom X to Z, and the wells correspond to $\tau\sim60^\circ, 180^\circ, 300^\circ$, respectively, corresponding to $\tau_2\approx180^\circ$ (H-O-C-Y), $\tau_1\approx180^\circ$ (H-O-C-Z), and $\tau_3\approx180^\circ$ (H-O-C-X), respectively. 

Tables \ref{tbl:HFCl}-\ref{tbl:FClI} present the vibrational energies and PV frequencies of the target molecules. In these tables, $\nu_A$ ($A$ = X, Y, Z = H, F, Cl, Br, I) represents the quantum number that corresponds to the well where the dihedral angle $A$-C-O-H is close to $180^{\circ}$ at the corresponding (approximate) $\tau$ values ($\tau\sim60^\circ,180^\circ,300^\circ$). ``-'' refers to mixed states that are delocalized into the three wells. 

The CXYZOH molecules can be broadly categorized into the follwoing two types based on their CXYZ units: those with hydrogen (H-containing) and those without hydrogen (H-non-containing). The H-containing molecules may be preferred from the measurement viewpoint because the deep well associated with the hydrogen vibrational frequency, denoted as $\nu_\mr{H}$, allows for high-energy transitions achievable by a CO$_2$ laser (around 1000 \cm). Contrarily, the H-non-containing molecules benefit from the localized characteristics of their torsional wavefunction, which arise from the proximity of the three local minimum depths. This feature would yield very long lifetimes owing to the predicted extremely narrow linewidth.

\par
%
%
\newcommand{\pvmol}{HFCl}
\begin{figure*}[!h]
  \begin{minipage}[b]{0.9\columnwidth}
    \centering
    \includegraphics[width=0.9\linewidth]{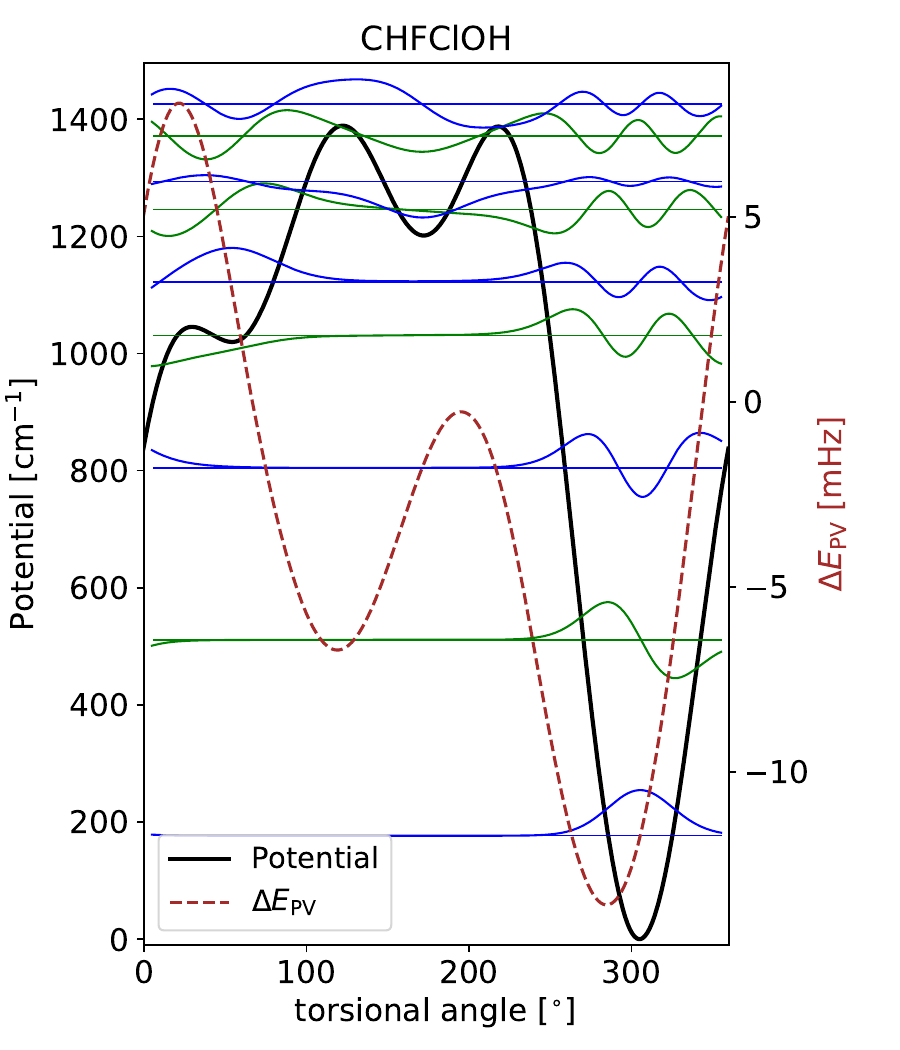}
        \caption{Visualization of PECs (black solid line, left axis), 1D torsional wavefunctions (blue and green solid lines, left axis), and \Epv\ curves (brown dotted line, right axis) of
        $R$-C{\pvmol}OH. 
        The \Epv\ at each torsional level is shown in Table 
        \ref{tbl:\pvmol}.}
    \label{fig:\pvmol}
  \end{minipage}
  \hspace{0.06\columnwidth} 
  \begin{minipage}[b]{0.9\columnwidth}
\newcommand{\pvmola}{HFBr}
    \centering
    \includegraphics[width=0.9\linewidth]{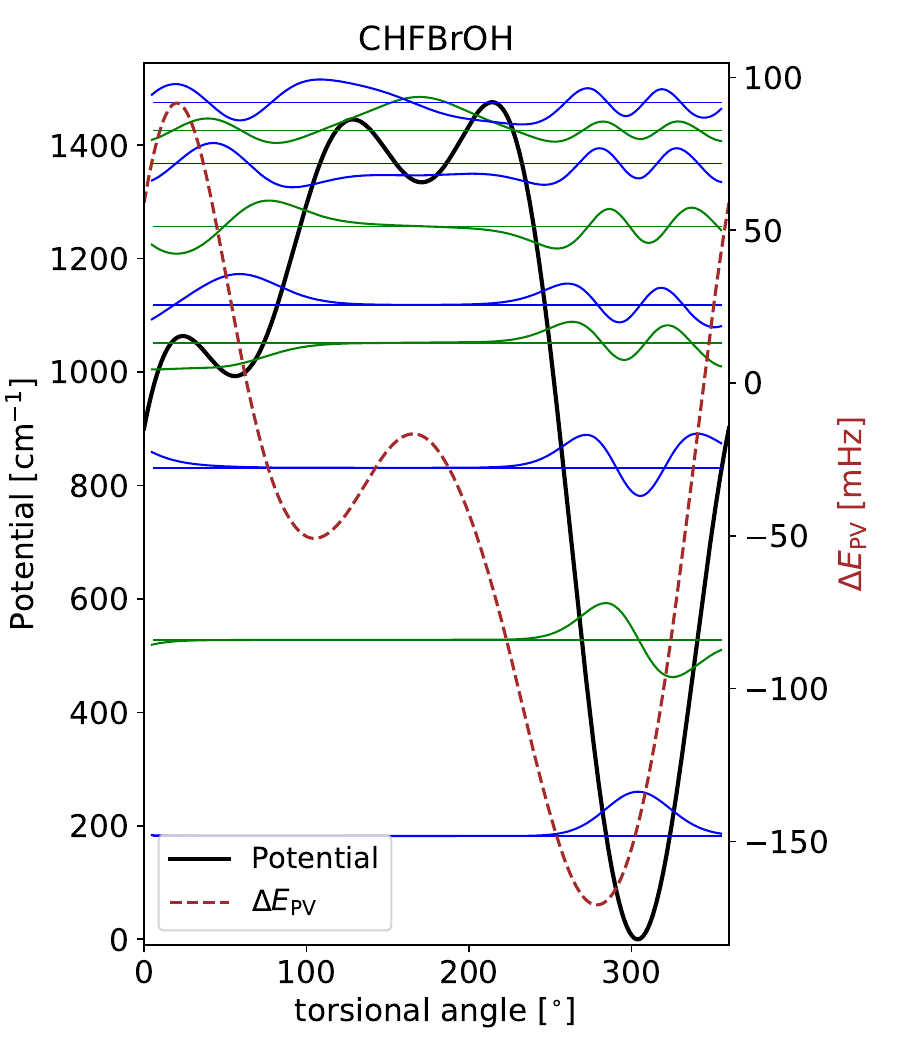}
        \caption{Visualization of PECs (black solid line, left axis), 1D torsional wavefunctions (blue and green solid lines, left axis), and \Epv\ curves (brown dotted line, right axis) of
        $R$-C{\pvmola}OH. 
        The \Epv\ at each torsional level is shown in Table 
        \ref{tbl:\pvmola}.}
    \label{fig:\pvmola}
      \end{minipage}
\end{figure*}

\begin{table*}[!h]
\begin{minipage}{8cm}
\caption{%
Vibrational energies ($\tilde{\nu}$ in \cm) referenced to the zero-point vibrational energy (\#1) and PV frequency shift (\delpv\ in mHz) of $R$-CHFClOH.
}
\label{tbl:HFCl}
\begin{tabular}{cccccr}
\hline
 & 60 & 180 & 300 & \multicolumn{2}{c}{CHFClOH}  \tabularnewline
\# &$\nu_\mr{F}$&$\nu_\mr{Cl}$&$\nu_\mr{H}$& $\tilde{\nu}$ & \delpv\tabularnewline
\hline
1 & 0 & 0 & 1 & 176.5 & $-$10.84\tabularnewline
2 & 0 & 0 & 2 & 334.5 & $-$8.77\tabularnewline
3 & 0 & 0 & 3 & 628.1 & $-$6.13\tabularnewline
4 & 0 & 0 & 4 & 854.3 & $-$0.88\tabularnewline
5 & 1 & 0 & 4 & 945.7 & $-$0.29\tabularnewline
6 & 1 & 0 & 5 & 1069.6 & $-$2.24\tabularnewline
7 & 0 & 1 & 0 & 1116.8 & $-$2.46\tabularnewline
8 & 2 & 1 & 5 & 1194.7 & $-$2.94\tabularnewline
9 & - & - & - & 1249.5 & $-$3.40\tabularnewline
10 & - & - & - & 1380.3 & $-$2.89\tabularnewline
\hline
\end{tabular}

\end{minipage}
\hfill
\begin{minipage}{8cm}
\caption{%
Vibrational energies ($\tilde{\nu}$ in \cm) referenced to the zero-point vibrational energy (\#1) and PV frequency shift (\delpv\ in mHz) of $R$-CHFBrOH.
}
\label{tbl:HFBr}
\begin{tabular}{ccccc r}
\hline
 & 60 & 180 & 300 & \multicolumn{2}{c}{CHFBrOH} \tabularnewline
\# &$\nu_\mr{F}$&$\nu_\mr{Br}$&$\nu_\mr{H}$& $\tilde{\nu}$ & \delpv\tabularnewline
\hline
1 & 0 & 0 & 1 & 182.0 & $-$135.31\tabularnewline
2 & 0 & 0 & 2 & 345.9 & $-$113.57\tabularnewline
3 & 0 & 0 & 3 & 649.3 & $-$83.71\tabularnewline
4 & 0 & 0 & 4 & 869.6 & 0.36\tabularnewline
5 & 1 & 0 & 4 & 936.0 & $-$28.90\tabularnewline
6 & 1 & 0 & 5 & 1075.1 & $-$27.02\tabularnewline
7 & 2 & 0 & 5 & 1185.4 & $-$43.66\tabularnewline
8 & 2 & 1 & 5 & 1244.0 & $-$27.36\tabularnewline
9 & - & - & - & 1292.9 & $-$46.38\tabularnewline
10 & - & - & - & 1427.5 & $-$38.58\tabularnewline
\hline
\end{tabular}
    \end{minipage}
\end{table*}


\clearpage
%
%
\section{Torsional quantum number for the CXYZOH molecules}
Traditionally, the quantum numbers at the vibrational ground states are defined as zero for all the vibrational modes. 
However, this convention would be insufficient in reflecting the character of the torsional wavefunction because `$\nu_{\tau}=0$' does not offer information on the bottom of the torsion potential at which the wavefunction is localized. Therefore, in this study, we defined the torsional quantum number as the number of the extremums of the vibrational wavefunction, except for the numerical noises, to clarify the character of the wavefunction.


\newcommand{\pvmolb}{HFI}
\begin{figure*}[!h]
\begin{minipage}[b]{0.9\columnwidth}
    \centering
    \includegraphics[width=0.9\linewidth]{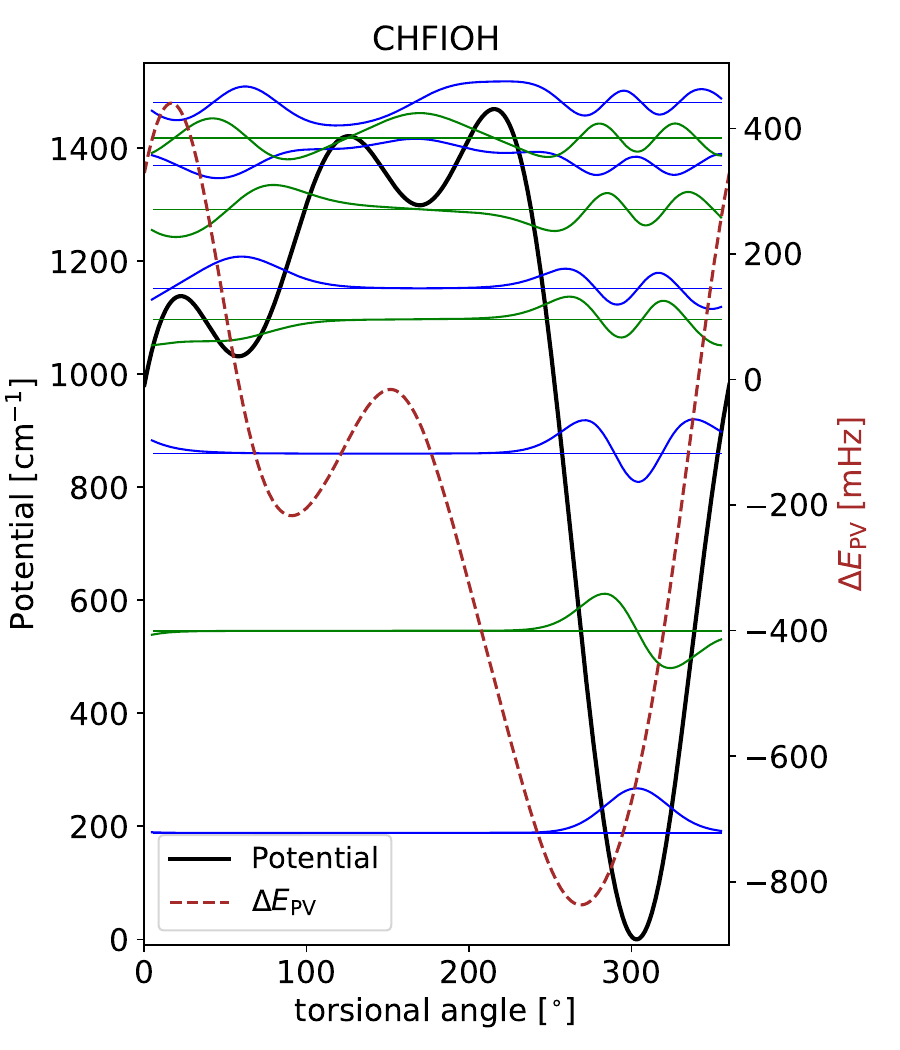}
        \caption{Visualization of PECs (black solid line, left axis), 1D torsional wavefunctions (blue and green solid lines, left axis), and \Epv\ curves (brown dotted line, right axis) of
        $R$-C{\pvmolb}OH. 
        The \Epv\ at each torsional level is shown in Table 
        \ref{tbl:\pvmolb}.}
    \label{fig:\pvmolb}
  \end{minipage}
  \hspace{0.06\columnwidth} 
  \begin{minipage}[b]{0.9\columnwidth}
\newcommand{\pvmolc}{HClBr}
    \centering
    \includegraphics[width=0.9\linewidth]{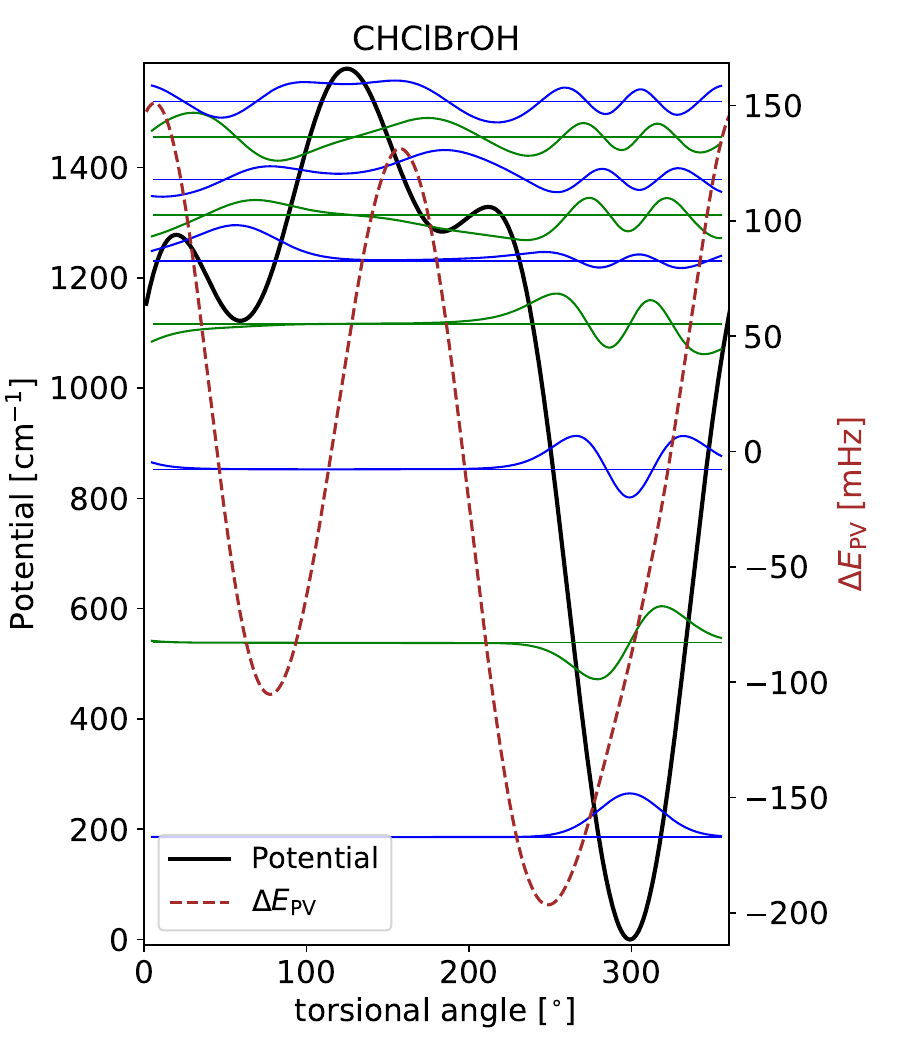}
        \caption{Visualization of PECs (black solid line, left axis), 1D torsional wavefunctions (blue and green solid lines, left axis), and \Epv\ curves (brown dotted line, right axis) of
        $R$-C{\pvmolc}OH. 
        The \Epv\ at each torsional level is shown in Table 
        \ref{tbl:\pvmolc}.}
    \label{fig:\pvmolc}
     \end{minipage}
\end{figure*}

\begin{table*}[!h]
 \begin{minipage}[t]{8cm}
 \caption{%
Vibrational energies ($\tilde{\nu}$ in \cm) referenced to the zero-point vibrational energy (\#1) and PV frequency shift (\delpv\ in mHz) of $R$-CHFIOH.
}
\label{tbl:HFI}
\begin{tabular}{ccccc r}
\hline
 & 60 & 180 & 300 & \multicolumn{2}{c}{CHFIOH} \tabularnewline
\# &$\nu_\mr{F}$&$\nu_\mr{I}$&$\nu_\mr{H}$& $\tilde{\nu}$ & \delpv\tabularnewline
\hline
1 & 0 & 0 & 1 & 188.2 & $-$603.57\tabularnewline
2 & 0 & 0 & 2 & 357.6 & $-$522.90\tabularnewline
3 & 0 & 0 & 3 & 671.3 & $-$406.44\tabularnewline
4 & 0 & 0 & 4 & 908.9 & $-$63.08\tabularnewline
5 & 1 & 0 & 4 & 963.0 & $-$154.39\tabularnewline
6 & 2 & 0 & 4 & 1103.2 & $-$125.74\tabularnewline
7 & 2 & 1 & 4 & 1181.1 & $-$166.44\tabularnewline
8 & 2 & 1 & 5 & 1229.7 & $-$172.83\tabularnewline
9 & - & - & - & 1292.5 & $-$225.94\tabularnewline
10 & - & - & - & 1422.9 & $-$181.43\tabularnewline
\hline
\end{tabular}
 \end{minipage}
    \hfill  
\begin{minipage}[t]{8cm} 
 \caption{%
Vibrational energies ($\tilde{\nu}$ in \cm) referenced to the zero-point vibrational energy (\#1) and PV frequency shift (\delpv\ in mHz) of $R$-CHClBrOH.
}
\label{tbl:HClBr}
\begin{tabular}{ccccc r}
\hline
 & 60 & 180 & 300 & \multicolumn{2}{c}{CHClBrOH} \tabularnewline
\# &$\nu_\mr{Cl}$&$\nu_\mr{Br}$&$\nu_\mr{H}$& $\tilde{\nu}$ & \delpv\tabularnewline
\hline
1 & 0 & 0 & 1 & 185.5 & $-$87.84\tabularnewline
2 & 0 & 0 & 2 & 352.7 & $-$80.38\tabularnewline
3 & 0 & 0 & 3 & 667.1 & $-$68.46\tabularnewline
4 & 0 & 0 & 4 & 931.1 & $-$47.54\tabularnewline
5 & 1 & 0 & 4 & 1045.2 & $-$32.28\tabularnewline
6 & 1 & 0 & 5 & 1128.2 & $-$43.21\tabularnewline
7 & 1 & 1 & 5 & 1192.8 & 28.58\tabularnewline
8 & 2 & 1 & 5 & 1269.7 & $-$11.72\tabularnewline
9 & - & - & - & 1334.0 & $-$1.29\tabularnewline
10 & - & - & - & 1458.2 & $-$0.47\tabularnewline
\hline
\end{tabular}
 \end{minipage}
\end{table*}

\clearpage

\newcommand{\pvmold}{HClI}
\begin{figure*}[!h]
\begin{minipage}[b]{0.9\columnwidth}
    \centering
    \includegraphics[width=0.9\linewidth]{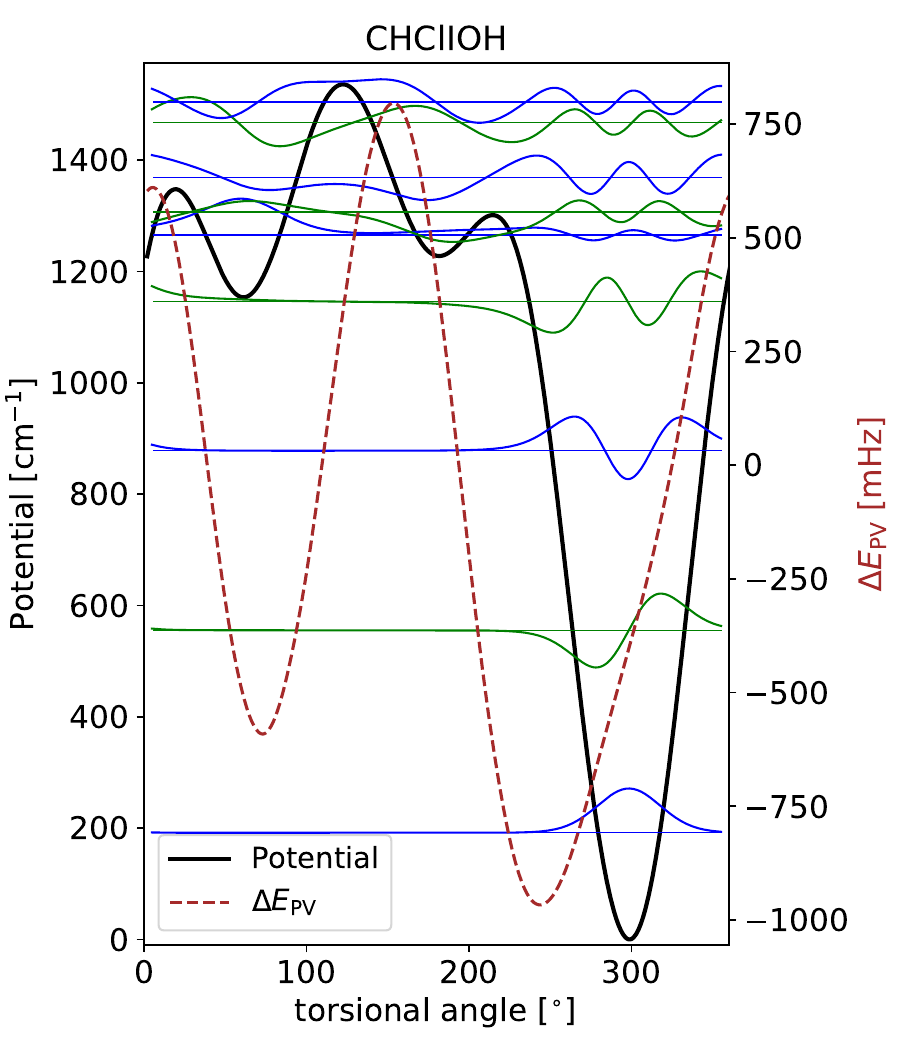}
        \caption{Visualization of PECs (black solid line, left axis), 1D torsional wavefunctions (blue and green solid lines, left axis), and \Epv\ curves (brown dotted line, right axis) of
        $R$-C{\pvmold}OH. 
        The \Epv\ at each torsional level is shown in Table 
        \ref{tbl:\pvmold}.}
    \label{fig:\pvmold}
  \end{minipage}
  \hspace{0.06\columnwidth} 
  \begin{minipage}[b]{0.9\columnwidth}
\newcommand{\pvmole}{FClI}
    \centering
    \includegraphics[width=0.9\linewidth]{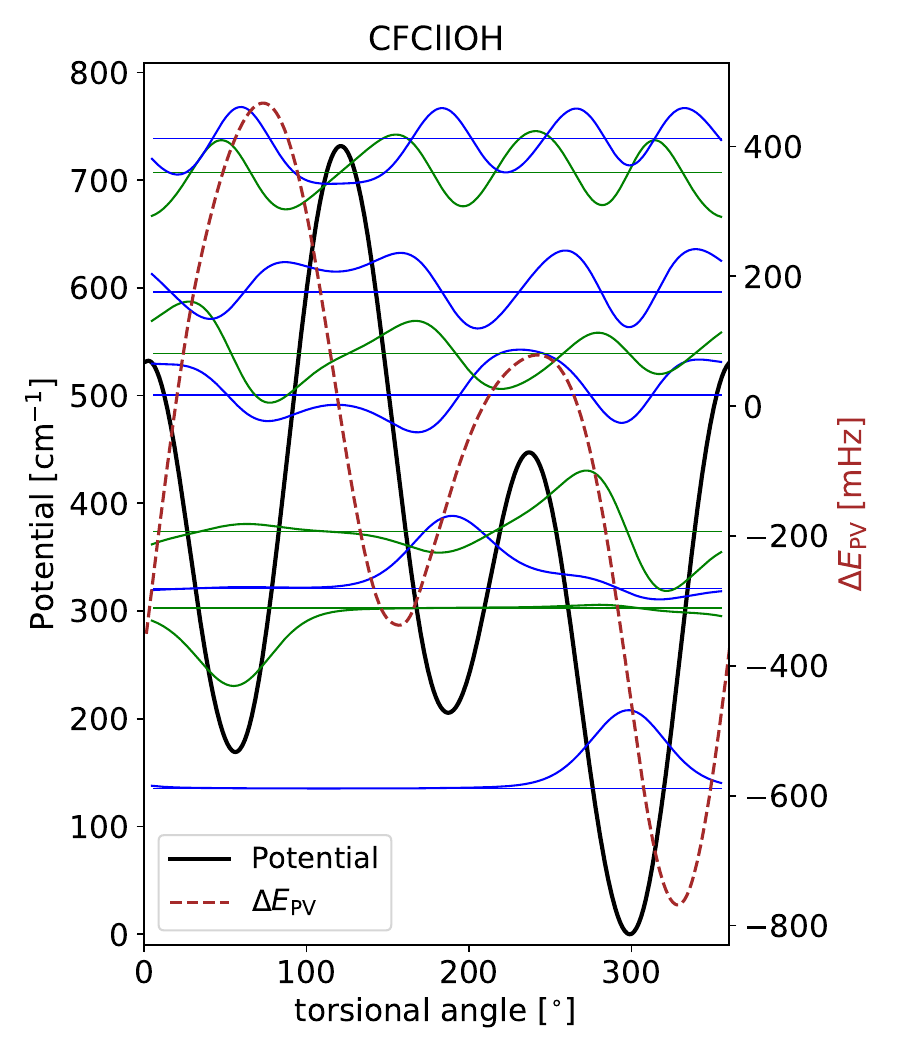}
        \caption{Visualization of PECs (black solid line, left axis), 1D torsional wavefunctions (blue and green solid lines, left axis), and \Epv\ curves (brown dotted line, right axis) of
        $S$-C{\pvmole}OH. 
        The \Epv\ at each torsional level is shown in Table 
        \ref{tbl:\pvmole}.}
    \label{fig:\pvmole}
    \end{minipage}
\end{figure*}

\begin{table*}[!h]
 \begin{minipage}[t]{8cm}
 \caption{%
Vibrational energies ($\tilde{\nu}$ in \cm) referenced to the zero-point vibrational energy (\#1) and PV frequency shift (\delpv\ in mHz) of $R$-CHClIOH.
}
\label{tbl:HClI}
\begin{tabular}{ccccc r}
\hline
 & 60 & 180 & 300 & \multicolumn{2}{c}{CHClIOH} \tabularnewline
\# &$\nu_\mr{Cl}$&$\nu_\mr{I}$&$\nu_\mr{H}$& $\tilde{\nu}$ & \delpv\tabularnewline
\hline
1 & 0 & 0 & 1 & 191.6 & $-$396.41\tabularnewline
2 & 0 & 0 & 2 & 363.7 & $-$384.84\tabularnewline
3 & 0 & 0 & 3 & 686.1 & $-$359.17\tabularnewline
4 & 0 & 0 & 4 & 954.3 & $-$316.16\tabularnewline
5 & 1 & 0 & 0 & 1073.5 & $-$305.51\tabularnewline
6 & 1 & 1 & 4 & 1115.2 & $-$48.16\tabularnewline
7 & 1 & 1 & 5 & 1177.1 & 46.05\tabularnewline
8 & 2 & 1 & 5 & 1275.4 & $-$96.26\tabularnewline
9 & 2 & 2 & 5 & 1312.6 & 61.83\tabularnewline
10 & - & - & - & 1448.7 & $-$13.87\tabularnewline
\hline
\end{tabular}
 \end{minipage}
    \hfill
\begin{minipage}[t]{8cm} 
 \caption{%
Vibrational energies ($\tilde{\nu}$ in \cm) referenced to the zero-point vibrational energy (\#1) and PV frequency shift (\delpv\ in mHz) of $S$-CFClIOH.
}
\label{tbl:FClI}
\begin{tabular}{ccccc r}
\hline
 & 60 & 180 & 300 & \multicolumn{2}{c}{CFClIOH} \tabularnewline
\# &$\nu_\mr{Cl}$&$\nu_\mr{I}$&$\nu_\mr{F}$& $\tilde{\nu}$ & \delpv\tabularnewline
\hline
1 & 0 & 0 & 1 & 135.3 & $-$412.04\tabularnewline
2 & 1 & 0 & 0 & 167.4 & 352.51\tabularnewline
3 & 0 & 1 & 0 & 185.6 & $-$126.32\tabularnewline
4 & 0 & 1 & 2 & 238.4 & $-$332.21\tabularnewline
5 & 2 & 2 & 2 & 365.2 & $-$126.10\tabularnewline
6 & - & - & - & 403.9 & 75.32\tabularnewline
7 & - & - & - & 460.7 & $-$169.53\tabularnewline
8 & - & - & - & 571.8 & $-$110.53\tabularnewline
9 & - & - & - & 603.5 & $-$47.60\tabularnewline
10 & - & - & - & 757.2 & $-$91.68\tabularnewline
\hline
\end{tabular}
\end{minipage}
\end{table*}


%
%
\section{Comparison with CAM-B3LYP* functional}
The enhancement mechanism is also examined using another DFT functional optimized for \Epv\ (the contribution only from Br and Cl atoms) in CHFClBr at the CCSD(T) level, which is noted as CAM-B3LYP*~\cite{Thierfelder2010PRA_CAMB3LYP*}.  

In the CAM-B3LYP approximation \cite{becke1993new,Stephens_JPC1994,Yanai2004CPL_CAMB4LYP}, electron-electron Coulomb operator into a short-range density functional and a long-range wave-function-based part is expressed by
\be
V_{\mathrm{ee}}=\sum_{i<j} \frac{1-\left[\alpha+\beta \operatorname{erf}\left(\mu r_{i j}\right)\right]}{r_{i j}}+\sum_{i<j} \frac{\alpha+\beta \operatorname{erf}\left(\mu r_{i j}\right)}{r_{i j}},
\ee
where $r_{i j}$ is the distance between two electrons $i$ and $j$, $\alpha=0.20$, $\beta= 0.12$, and $µ = 0.90$ are the parameters optimized in Ref.~\citenum{Thierfelder2010PRA_CAMB3LYP*}.

The comparison between the CAM-B3LYP* and PBE0 functionals is shown in Table \ref{tbl:CAM_PEC}. The maximum deviation is --0.12 mHz ($\tau=306^\circ$), which is much smaller than the maximum absolute \Epv\ value shown at $\tau=288^\circ$, --6.86 (--6.77) mHz at the CAM-B3LYP* (PBE0) functionals. 
The maximum difference in the PEC is relatively large (--57.6 \cm\ at $\tau=234^\circ$), but the torsional angles that show the local minimum and maximum are very similar. 

Table \ref{tbl:CAM_1D} lists the PV
frequencies and vibrational energies calculated with the CAM-B3LYP* and PBE0 functionals. The relative difference of \delpv\ becomes large when its absolute value of \delpv\ is close to zero (e.g., \#4 and \#5). However, the differences ($\Delta$) are smaller than the most sensitive transition to the PV (between \#1 and \#5, $\sim 11$ mHz). In the states in which the vibrational wavefunction is plane-wave-like (i.e., high-energy vibrational states), the values of \delpv\ at the CAM-B3LYP* and PBE0 levels are very close. From this and the comparison shown in Table \ref{tbl:CAM_PEC}, the difference between CAM-B3LYP* and PBE0 would be mainly due to the vibrational wavefunction. Although the benchmark data is not reported for the CAM-B3LYP*, the validity of the PBE0 function for the torsional PEC is assessed in Ref \cite{Tahchieva2018JCTC_tor_DFT}. 
%
%
\begin{table*}[!h]
\caption{%
\Epv\ and PEC of CHFClOH obtained with CAM-B3LYP* and PBE0 functionals based on the eAMF-X2C Hamiltonian (with the Gaunt term) as a function of torsional angles. The dyall.3zp basis set with tight exponents is used. The differences between the two functionals ($\Delta$) are also listed. The geometries were optimized at the PBE1PBE/Def2TZVPP level.
}
\label{tbl:CAM_PEC}
\begin{tabular}{c SSSSSS}
\hline\hline
 & \multicolumn{3}{c}{\Epv\ (mHz)} & \multicolumn{3}{c}{PEC (\cm)}\tabularnewline
$\tau$ & {CAM-B3LYP{*}} & {PBE0} & {$\Delta$} & {CAM-B3LYP{*}} & {PBE0} & {$\Delta$}\tabularnewline
\hline
0 & 2.50 & 2.51 & -0.01 & 865.2 & 837.5 & 27.7\tabularnewline
9 & 3.51 & 3.50 & 0.01 & 979.9 & 953.2 & 26.7\tabularnewline
18 & 4.03 & 4.00 & 0.03 & 1044.4 & 1020.0 & 24.3\tabularnewline
27 & 3.99 & 3.95 & 0.04 & 1066.1 & 1044.4 & 21.7\tabularnewline
36 & 3.49 & 3.45 & 0.04 & 1058.7 & 1040.9 & 17.8\tabularnewline
45 & 2.63 & 2.59 & 0.04 & 1042.4 & 1027.7 & 14.7\tabularnewline
54 & 1.58 & 1.55 & 0.03 & 1031.8 & 1019.7 & 12.1\tabularnewline
63 & 0.49 & 0.47 & 0.02 & 1039.9 & 1030.6 & 9.3\tabularnewline
72 & -0.52 & -0.53 & 0.01 & 1074.8 & 1068.6 & 6.2\tabularnewline
81 & -1.44 & -1.43 & -0.01 & 1131.9 & 1128.1 & 3.8\tabularnewline
90 & -2.20 & -2.18 & -0.02 & 1204.0 & 1202.6 & 1.4\tabularnewline
99 & -2.80 & -2.77 & -0.03 & 1278.5 & 1279.5 & -1.0\tabularnewline
108 & -3.20 & -3.16 & -0.04 & 1340.2 & 1343.3 & -3.2\tabularnewline
117 & -3.38 & -3.34 & -0.04 & 1376.3 & 1381.6 & -5.3\tabularnewline
126 & -3.34 & -3.30 & -0.04 & 1379.9 & 1387.3 & -7.4\tabularnewline
135 & -3.08 & -3.05 & -0.03 & 1352.8 & 1362.2 & -9.4\tabularnewline
144 & -2.63 & -2.61 & -0.02 & 1304.2 & 1315.7 & -11.5\tabularnewline
153 & -2.06 & -2.06 & 0.00 & 1249.0 & 1263.2 & -14.2\tabularnewline
162 & -1.45 & -1.47 & 0.02 & 1203.1 & 1220.8 & -17.7\tabularnewline
171 & -0.87 & -0.91 & 0.04 & 1179.5 & 1201.7 & -22.2\tabularnewline
180 & -0.41 & -0.47 & 0.06 & 1184.3 & 1212.3 & -27.9\tabularnewline
189 & -0.11 & -0.19 & 0.08 & 1216.1 & 1250.4 & -34.3\tabularnewline
198 & -0.06 & -0.15 & 0.09 & 1263.2 & 1305.2 & -42.0\tabularnewline
207 & -0.28 & -0.37 & 0.09 & 1309.7 & 1358.1 & -48.4\tabularnewline
216 & -0.80 & -0.88 & 0.08 & 1332.8 & 1386.8 & -54.0\tabularnewline
225 & -1.61 & -1.68 & 0.07 & 1312.6 & 1369.8 & -57.2\tabularnewline
234 & -2.63 & -2.68 & 0.05 & 1234.1 & 1291.7 & -57.6\tabularnewline
243 & -3.78 & -3.81 & 0.03 & 1093.3 & 1149.6 & -56.3\tabularnewline
252 & -4.87 & -4.87 & 0.00 & 902.9 & 954.3 & -51.4\tabularnewline
261 & -5.79 & -5.77 & -0.03 & 683.3 & 727.6 & -44.4\tabularnewline
270 & -6.47 & -6.41 & -0.05 & 460.5 & 496.4 & -35.9\tabularnewline
279 & -6.83 & -6.75 & -0.08 & 262.9 & 289.4 & -26.6\tabularnewline
288 & -6.86 & -6.77 & -0.10 & 111.0 & 128.0 & -17.0\tabularnewline
297 & -6.56 & -6.45 & -0.11 & 20.2 & 27.5 & -7.3\tabularnewline
306 & -5.91 & -5.80 & -0.12 & 2.4 & 0.7 & 1.7\tabularnewline
315 & -4.93 & -4.82 & -0.11 & 55.3 & 45.6 & 9.7\tabularnewline
324 & -3.66 & -3.56 & -0.10 & 170.4 & 153.5 & 17.0\tabularnewline
333 & -2.15 & -2.07 & -0.09 & 332.8 & 310.3 & 22.5\tabularnewline
342 & -0.51 & -0.45 & -0.06 & 519.8 & 493.8 & 26.0\tabularnewline
351 & 1.10 & 1.13 & -0.03 & 705.7 & 678.1 & 27.6\tabularnewline
360 & 2.50 & 2.51 & -0.01 & 865.2 & 837.5 & 27.7\tabularnewline
\hline\hline
\end{tabular}
\end{table*}

%
%
\begin{table*}[!h]
\caption{%
Vibrational energies ($\tilde{\nu}$ in \cm) referenced to the zero-point vibrational energy (\#1) and PV frequency shift (\delpv\ in mHz) of $R$-CHFClOH. The \Epv\ curve and PEC shown in Table \ref{tbl:CAM_PEC} are used for the vibrational computations. The states are assigned referring to FIG. \ref{fig:HFCl} .
}
\label{tbl:CAM_1D}
\begin{tabular}{cccc SSSSSS}
\hline\hline
 & 60 & 180 & 300 & \multicolumn{3}{c}{$\tilde{\nu}$ (\cm)} & \multicolumn{3}{c}{\delpv\ (mHz)}\tabularnewline
 & F & Cl & H & {CAM-B3LYP{*}} & {PBE0} & {$\Delta$} & {CAM-B3LYP{*}} & {PBE0} & {$\Delta$}\tabularnewline
 \hline
1 & 0 & 0 & 1 & 176.0 & 176.5 & -0.6 & -11.25 & -10.84 & -0.41 \tabularnewline
2 & 0 & 0 & 2 & 333.4 & 334.5 & -1.1 & -9.18 & -8.77 & -0.41 \tabularnewline
3 & 0 & 0 & 3 & 626.6 & 628.1 & -1.6 & -6.60 & -6.13 & -0.47 \tabularnewline
4 & 0 & 0 & 4 & 857.6 & 854.3 & 3.3 & -1.75 & -0.88 & -0.88 \tabularnewline
5 & 1 & 0 & 4 & 950.6 & 945.7 & 4.9 & 0.28 & -0.29 & 0.57 \tabularnewline
6 & 1 & 0 & 5 & 1065.0 & 1069.6 & -4.6 & -2.22 & -2.24 & 0.02 \tabularnewline
7 & 0 & 1 & 0 & 1094.2 & 1116.8 & -22.7 & -2.18 & -2.46 & 0.28 \tabularnewline
8 & 2 & 1 & 5 & 1187.1 & 1194.7 & -7.6 & -2.75 & -2.94 & 0.19 \tabularnewline
9 & - & - & - & 1231.5 & 1249.5 & -18.0 & -3.37 & -3.40 & 0.03 \tabularnewline
10 & - & - & - & 1368.8 & 1380.3 & -11.5 & -2.84 & -2.89 & 0.06 \tabularnewline
11 & - & - & - & 1382.0 & 1395.7 & -13.7 & -3.18 & -3.20 & 0.02 \tabularnewline
12 & - & - & - & 1597.2 & 1608.5 & -11.3 & -2.98 & -3.04 & 0.06 \tabularnewline
13 & - & - & - & 1598.4 & 1609.9 & -11.5 & -3.03 & -3.04 & 0.02 \tabularnewline
14 & - & - & - & 1870.7 & 1881.4 & -10.7 & -3.05 & -3.08 & 0.03 \tabularnewline
15 & - & - & - & 1870.8 & 1881.6 & -10.8 & -3.04 & -3.07 & 0.03 \tabularnewline
16 & - & - & - & 2191.2 & 2201.5 & -10.4 & -3.09 & -3.12 & 0.02 \tabularnewline
17 & - & - & - & 2191.2 & 2201.6 & -10.4 & -3.09 & -3.11 & 0.02 \tabularnewline
18 & - & - & - & 2557.6 & 2567.7 & -10.1 & -3.14 & -3.15 & 0.02 \tabularnewline
19 & - & - & - & 2557.7 & 2567.8 & -10.1 & -3.13 & -3.15 & 0.02 \tabularnewline
20 & - & - & - & 2969.2 & 2979.2 & -9.9 & -3.17 & -3.18 & 0.01 \tabularnewline
\hline\hline
\end{tabular}
\end{table*}

\end{document}